\documentclass[a4paper,preprint]{emulateapj}
\usepackage{amstext}
\usepackage{apjfonts}
\usepackage{amsmath}
\usepackage{psfig}

\begin{document}
\title{An Off-Axis Model for GRB 031203}

\author{Enrico Ramirez-Ruiz\altaffilmark{1,2,3}, Jonathan
Granot\altaffilmark{1,4}, Chryssa Kouveliotou\altaffilmark{1,5}, S.
E. Woosley\altaffilmark{1,6}, Sandy K. Patel\altaffilmark{1,7}, and
Paolo A. Mazzali\altaffilmark{1,8}} \altaffiltext{1}{Institute for
Nuclear Theory, University of Washington, Seattle, Washington
98195-1550, USA} \altaffiltext{2}{Institute for Advanced Study,
Einstein Drive, Princeton, NJ 08540, USA} \altaffiltext{3}{Chandra
Fellow} \altaffiltext{4}{KIPAC, Stanford University, P.O. Box 20450,
MS 29, Stanford, CA 94309, USA} \altaffiltext{5}{NASA/Marshall Space
Flight Center, NSSTC, XD-12, 320 Sparkman Dr., Huntsville, AL 35805,
USA} \altaffiltext{6}{Department of Astronomy and Astrophysics,
University of California, Santa Cruz, CA 95064, USA}
\altaffiltext{7}{USRA, NSSTC, SD-50, 320 Sparkman Dr., Huntsville, AL
35805, USA} \altaffiltext{8}{INAF Osservatorio Astronomico, Via
Tiepolo, 11, 34131 Trieste, Italy}

\begin{abstract} 

The low luminosity radio emission of the unusually faint GRB 031203
has been argued to support the idea of a class of intrinsically
sub-energetic gamma-ray bursts (GRBs), currently comprising two
members. While low energy GRBs probably exist, we show that the
collective prompt and multiwavelength observations of the afterglow of
GRB 031203 do not necessarily require a sub-energetic nature for that
event. In fact, the data are more consistent with a typical, powerful
GRB seen at an angle of about twice the opening angle of the central
jet. The {\it intrinsic} peak energy, $E_{\rm p}$, of GRB~031203 then
becomes $\sim 2\;$MeV, similar to many other GRBs.

\end{abstract}


\section{Introduction}

The first evidence that gamma-ray bursts might have a broad range of
energies came with the discovery of GRB 980425, the first GRB also to
be associated with a Type Ib/c supernova, SN~1998bw \citep{Galama98}.
While unremarkable in its time scale and spectrum, GRB 980425 had a
total gamma-ray energy, assuming isotropic emission, of only
$E_{\gamma,{\rm iso}} \sim 10^{48}\;$ergs, some 4-6 orders of
magnitude less than a typical GRB \citep{Frail01,GGL04}.  Significant
interest was aroused at the time by the possibility that such
lower-energy bursts might be more common than had been thought, but
hard to detect given the current instrumental sensitivities. It took
five more years before another event, GRB 031203, provided additional
support for a faint population of GRBs. At a cosmological distance of
$z = 0.1055$ \citep{Prochaska04}, GRB 031203 was also atypical in its
gamma-ray budget with $E_{\gamma,{\rm iso}} \sim 10^{50}\;$ergs
\citep{SLS04}.  In fact its gamma-ray power was intermediate between
GRB 980425 and more typical bursts with (isotropic) energies of
$10^{52}-10^{54}\;$ergs (Frail et al. 2001; Ghirlanda et al. 2004).
The burst profile was smooth and similar to GRB 980425, consisting of
a single peak lasting about 20 s and a peak energy above $190\;$keV
\citep{SLS04}.

Soon afterwards, an optical counterpart was identified and follow-up
observations by several telescopes revealed a supernova, SN 2003lw,
with a spectrum very similar to that of SN 1998bw
\citep{Malesani04,Thomsen04,galyam04,cobb04}.  Subsequent X-ray
observations of GRB 031203 with {\it XMM} and {\it Chandra} identified
an X-ray source coincident with the optical transient. The decline
rate and the isotropic luminosity of the X-ray afterglow also ranked
the event as intermediate between GRB 980425 and classical GRBs
\citep{Kouveliotou04}. A very faint counterpart was also detected at
centimeter wavelengths where it displayed a peak luminosity more than
two orders of magnitude fainter than typical radio afterglows
\citep{Frail03}, but again comparable to that of GRB 980425
\citep{Kulkarni98}.

Given the many similarities with GRB 980425, it has been argued
\citep[][hereafter S04; Sazonov et al. 2004]{Soderberg04} that the
{\it only} explanation of the faint nature of both GRB~031203 and
GRB~980425 is that they were intrinsically sub-energetic, that is the
energy ejected in relativistic matter at all angles was orders of
magnitude less than in all other GRBs studied to date. Further it has
been suggested that the afterglow data are only consistent with a
nearly spherical explosion - that GRB 031203 was not a jet-like
phenomenon (S04; Sazonov et al. 2004). We disagree with both
conclusions and show here that the data of GRB 031203, especially the
early X-ray afterglow light curve, do not require a sub-energetic
nature for this event, and are in fact more consistent with a model in
which GRB 031203 was a typical, powerful {\it jetted} GRB viewed
off-axis.\footnote{The reader is referred to \citet{GR-RP05} and
references therein for a detailed analysis of the off-axis model in
relation to X-ray flashes and X-ray rich GRBs.}

\section{Calculation of Afterglow Emission} 
\label{model}
The afterglow light curves presented here are calculated using model 1
of \citet{GK03}. The deceleration of the flow is calculated from the
mass and energy conservation equations and the energy per solid angle
$\epsilon$ is taken to be independent of time. The local emissivity is
calculated using the conventional assumptions of synchrotron emission
from relativistic electrons that are accelerated behind the shock into
a power-law distribution of energies, $N(\gamma_e) \propto
\gamma_e^{-p}$ for $\gamma_e > \gamma_m$, where the electrons and the
magnetic field hold fractions $\epsilon_e$ and $\epsilon_B$,
respectively, of the internal energy of the shocked fluid.  
The synchrotron spectrum is
taken to be a piecewise power law \citep{SPN98}. In \S \ref{X-ray} we
begin with a simple model where we assume that the outflow is
spherical. More realistic jet models are considered in \S
\ref{offaxis} where the Lorentz factor $\gamma$ and $\epsilon$ are
assumed, within the jet aperture, to be independent of the angle
$\theta$ as measured from the jet axis.  The lateral spreading of the
jet is neglected. This approximation is consistent with results of
numerical simulations \citep{Granot01} which show relatively little
lateral expansion as long as the jet is relativistic. The light curves
for observers located at different angles, $\theta_{\rm obs}$, with
respect to the jet axis are calculated by applying the appropriate
relativistic transformation of the radiation field from the local rest
frame of the emitting fluid to the observer frame and integrating
over equal photon arrival time surfaces \citep{Granot02,R-RM04}.

\section{The Importance of the X-Ray Light Curve}
\label{X-ray}

GRB 031203, or at least its gamma-rays directed at us, was certainly
very weak. A straightforward interpretation might be that the GRB was
deficient in all its emissions in all directions (S04). This idea is
compatible with the afterglow light curve at radio frequencies.
However, when one combines the fact that a 20 s long GRB was observed,
as well as an X-ray and infra-red afterglow, the situation is more
constrained.

\begin{figure}
\plotone{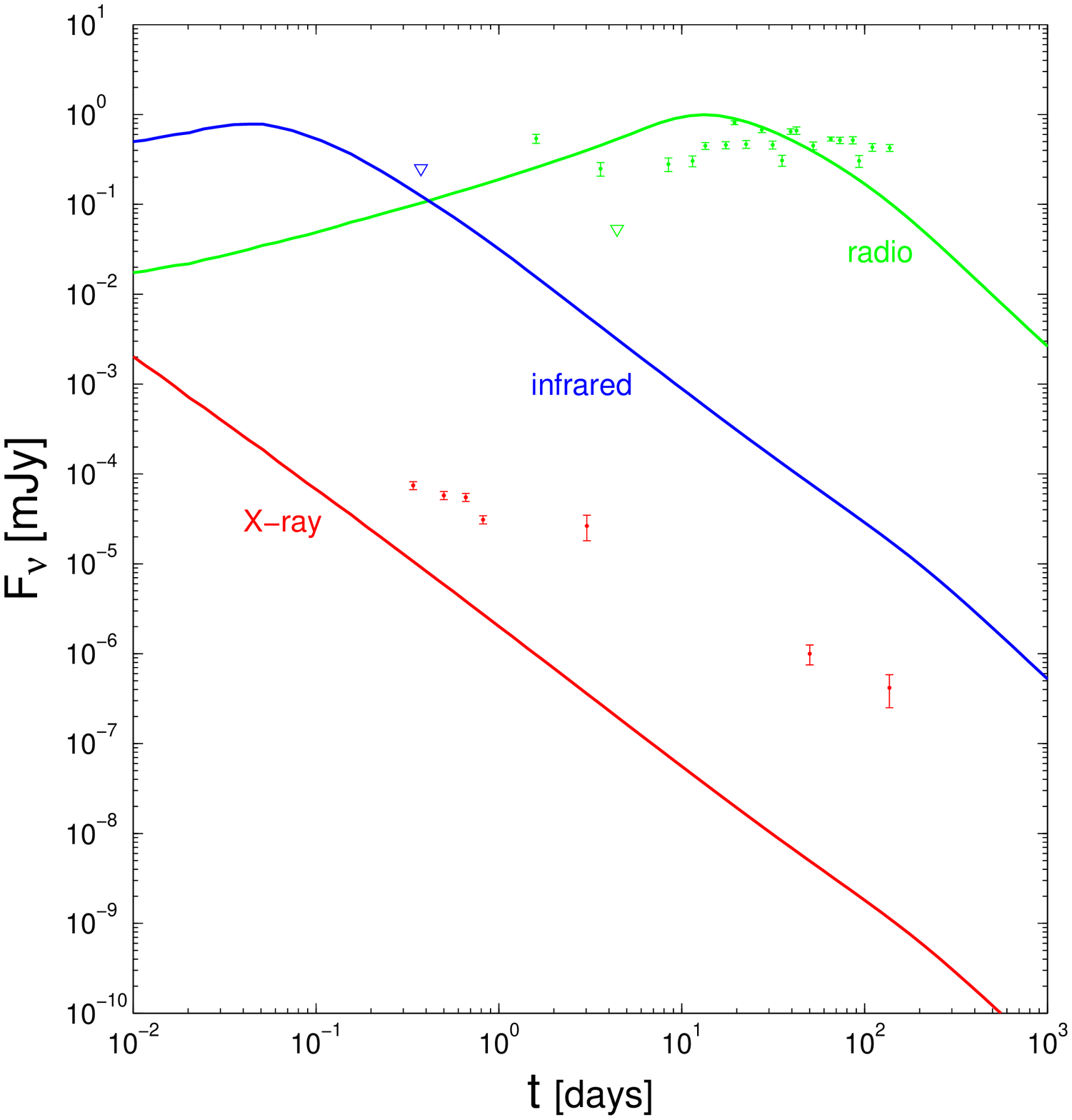}
\caption{{\footnotesize Afterglow emission from a spherical,
sub-energetic blast-wave. A tentative fit to the radio (8.5 GHz; S04),
infrared (K-band; Malesani et al. 2004), and X-ray (0.3-10 keV; Watson
et al. 2004b).The micro-physical parameters and the properties of both
the external medium and burst energetics are chosen to exactly match
those derived by S04 for the emission of GRB 031203. The last X-ray
point was obtained with ~30 ks of Director's Discretionary Time of
Chandra. During that observation we detected a source with flux of
$4\pm3 \times10^{-15}$ erg cm$^{-2}$ s$^{-1}$, assuming a power law
photon index of 1.7 and $N_{\rm H}$ consistent with the previous
Chandra and XMM observations.}}
\label{fig1}
\end{figure}

The resulting lightcurves for a sub-energetic spherical model are
plotted against the data in Fig.  \ref{fig1}. The model parameters are
chosen to coincide with those of S04: an energy of $E=1.7\times
10^{49}\;$ergs, a uniform external medium of number density
$n=0.6\;{\rm cm}^{-3}$, $p=2.6$, $\epsilon_e=0.4$ and
$\epsilon_B=0.2$.  Even though the model fits moderately well the
radio and infrared light curves (given the sparse data for the
latter), it is inconsistent with the slow decline of the X-ray light
curve during the first 100 days. The following point should be
emphasized here. The dynamical model used here is different from that
used by S04. This explains why our fit to the radio data is slightly
poorer in quality despite using similar model parameters. A similar
goodness of fit to the radio lightcurve can be easily achieved by
iterating over the physical parameters.  Such an exercise, however,
cannot at the same time provide an acceptable fit to the X-ray light
curve. In fact, we find that most spherical models underpredict the
late time X-ray flux by at least two orders of magnitude and cannot
account for the slow initial decline rate seen in the X-ray afterglow,
$F_\nu \propto t^{-\alpha}$, with $\alpha \approx 1/4$. This argues
against a spherical explosion with low energy content.

It might be possible, for instance, that in addition to the quasi
spherical, relativistic component (relevant to the afterglow) there is
also a subrelativistic outflow with lower $\gamma$ (heavier loading of
baryons) ejected by SN 2003lw in other directions. This slower matter
could in principle produce a nearly flat X-ray light curve for the
first few days, followed by a decay as the matter decelerated in the
stellar wind \citep{Waxman04,GR04}. This type of behavior bears some
similarities to the X-ray light curve seen in GRB 980425
\citep{Kouveliotou04}. This modified geometry, however, could not meet
the constraints posed by the observations. This is because the
corresponding (shock driven) radio emission produced by SN 2003lw
would be $\sim 30$ times too high, thus rendering this type of model
unacceptable.

\section{An Off-Axis Model}
\label{offaxis}

\begin{figure}
\plotone{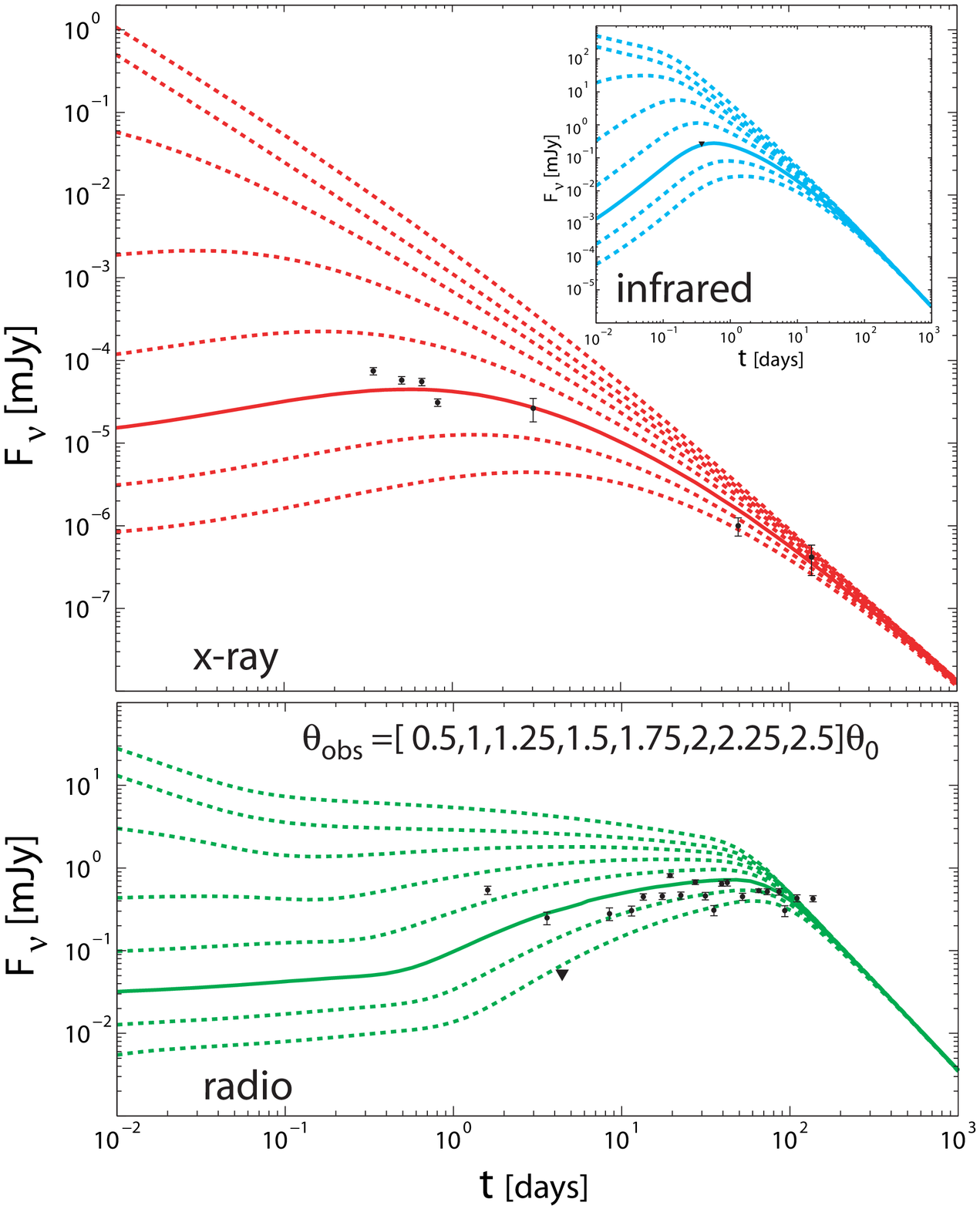}
\caption{{\footnotesize Afterglow emission from a sharp edged uniform jet in GRB
  031203. Light curves calculated for various viewing angles
  $\theta_{\rm obs}$ for a GRB with the standard parameters $E_{\rm
    jet}=3 \times 10^{50}$ erg, $p = 2.4$, $\epsilon_e = 0.15$,
  $\epsilon_B = 0.02$, $\theta_0 = 5^\circ$, and
  $A_*=(\dot{M}/10^{-5}\,{\rm M_\odot\,yr^{-1}})(v_w/ 10^{3}\,{\rm
    km\,s^{-1}})^{-1}$=0.1. The data for GRB 031203 can be reasonably
  fit by different sets of model parameters (i.e. the parameters
  cannot be uniquely determined by the data). For example, a
  sharp-edged jet with $\theta_0 = 3.5^\circ$ seen at $\theta_{\rm
    obs}\approx 2.25\theta_0$ gives also a reasonably good description
  of the observations provided that $\epsilon_e = 0.1$ and $\epsilon_B
  = 0.04$.}}
\label{fig2}
\end{figure}

\begin{figure*}
\plotone{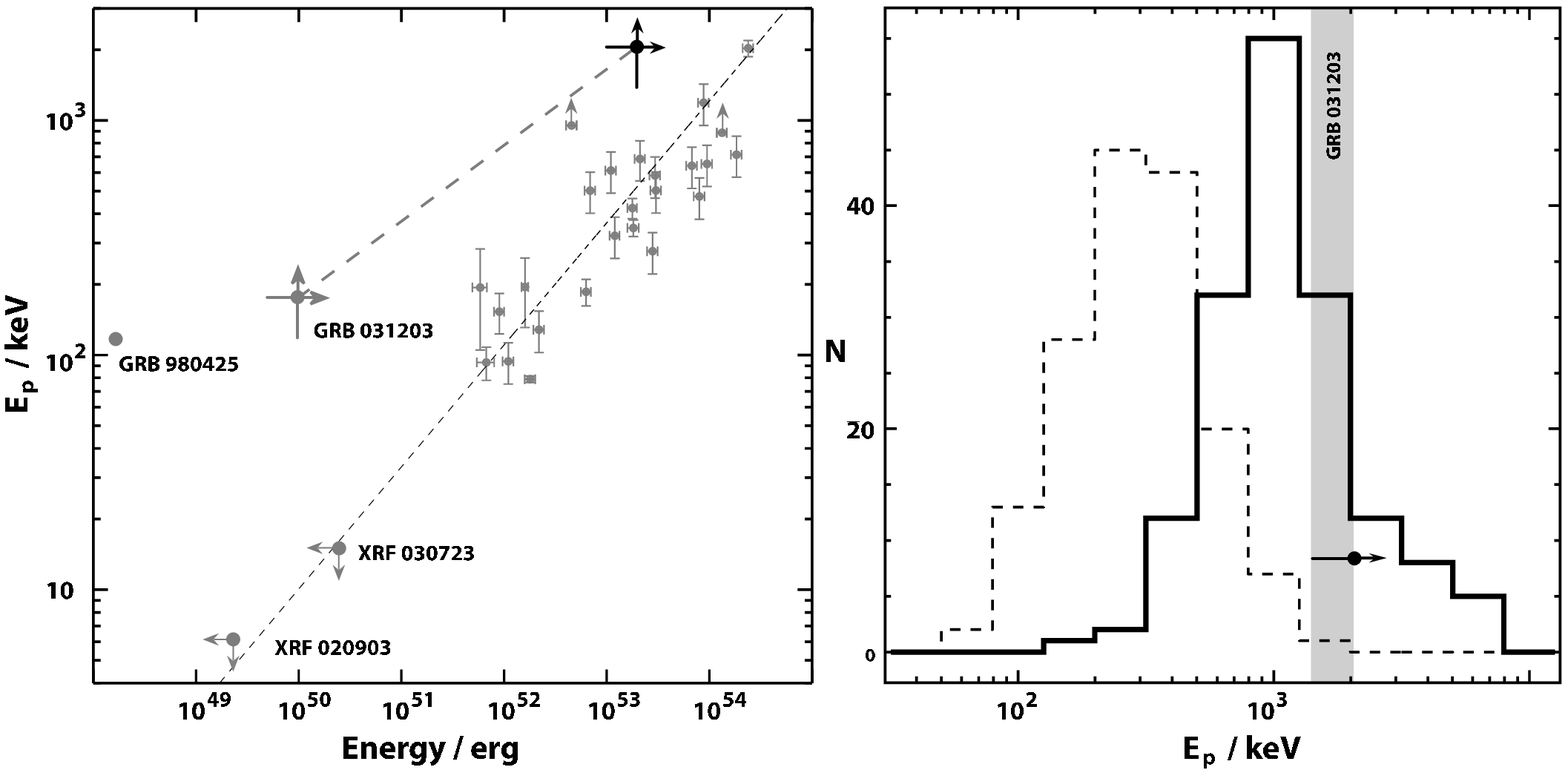}
\caption{{\footnotesize  Constraints on the possible existence of a misaligned,
  sharp-edged jet in GRB 031203. {\it Left Panel:} The location of GRB
  031203 in the $E_{\rm p} - E_{\gamma,{\rm iso}}$ plane. The
  compilation of observed $E_{\rm p}$ and $E_{\gamma,{\rm iso}}$ in
  the source frame derived by \citet{GGL04} are also illustrated.  If
  GRB 031203 was viewed on-axis (at $\theta_{\rm obs} < \theta_0$),
  the peak of the spectrum and the isotropic equivalent energy would
  be $\sim 2\;$MeV and $\sim 10^{53}\;$ergs, respectively (black
  symbol). {\it Right Panel:} Histogram of burst peak energies in
  their cosmological rest frame for BATSE events \citep{L-RR-R02}.
  Superposed on the plot (dotted line) is the histogram of the
  observed peak energy.}}
\label{fig3}
\end{figure*}

Given that most GRBs are collimated into narrow jets \citep{Frail01},
their observed properties will inevitably vary depending upon the
angle, $\theta_{\rm obs}$, from their symmetry axis at which they are
viewed. If we assume a homogeneous sharp-edged jet, the burst seen by
all observers located within the initial jet aperture, $\theta_{\rm
obs} < \theta_0$, is practically the same, but beyond the edges of the
jet the emission declines precipitously
\citep{WL99,Granot02,YIN02}. In the latter case, the observed prompt
GRB emission and its early afterglow are very weak, owing to the
relativistic beaming of photons away from the line of sight. Thus, an
observer at $\theta_{\rm obs} > \theta_0$ sees a rising afterglow
light curve at early times (as the Lorentz factor decreases with time)
peaking when the jet Lorentz factor reaches $\sim 1/(\theta_{\rm
obs}-\theta_0)$ and approaching that seen by an on-axis observer at
later times. This is because the emission remains at a very low level
until the Doppler cone of the beam intersects the observer's line of
sight. This can be seen by comparing the $\theta_{\rm obs}=\theta_0$
and $\theta_{\rm obs}=2\theta_0$ curves in Fig. \ref{fig2}.

The off-axis jet interpretation for GRB 031203 requires the viewing 
angle to have been $\theta_{\rm obs}\sim 2\theta_0$ (Fig.  \ref{fig2}).  
This interpretation assumes that our line of sight is a few degrees 
from a sharp-edged conical jet. A misaligned jet with a typical energy
expanding into a stellar wind with properties similar to those of
Wolf-Rayet stars is consistent with the observations, especially with
the slow initial decline rates seen in both the X-ray
\citep{Watson04a} and radio (S04) afterglow\footnote{When comparing
model predictions with radio observations one should expect an
approximate -- rather than exact -- agreement, as large fluctuations
seen in the centimeter-wave radio fluxes are likely due to
interstellar scintillation when the early fireball is nearly a point
source.}.  Interestingly, if the jet axis had been closer to the
observer's direction ($\theta_{\rm obs} < 2 \theta_0$), the brightness
of its infrared afterglow would have prevented the detection of SN
2003lw \citep{Malesani04}.

The constraints imposed by the properties of the afterglow data thus
favor the idea that GRB 031203 was a {\it typical} GRB jet seen at
$\theta_{\rm obs} > \theta_0$.  One question that naturally arises is
whether the observed gamma-ray flux of GRB 031203 can be explained
within the framework of this model. We consider below a geometry of a
jet with sharp edges seen at $\theta_{\rm obs} > \theta_0$; in that
case, the prompt emission comes from narrowly beamed material moving
along the edge of the jet which is closest to our line of sight.  This
is since the relativistic beaming of light away from our line of sight
is smallest within this region when compared to other parts of the
jet.

Because of the relativistic motion of jet ejecta, with Lorentz factor
$\gamma \gtrsim 100$ during gamma-ray emission, the gamma-rays are
concentrated into a cone of opening angle comparable to the jet
opening angle $\theta_0$ (assuming $\theta_0 > 1/\gamma$). Thus, if
the jet is viewed from a direction making an angle larger than
$\theta_0$ with the jet axis, the gamma-ray flux may be strongly
suppressed. For an off-axis GRB jet with bulk Lorentz factor $\gamma$,
$E_{\gamma,{\rm iso}} \propto [\gamma(\theta_{\rm obs}
-\theta_0)]^{-6}$ (for $\theta_{\rm obs}-\theta_0\gtrsim 1/\gamma$),
while the typical peak photon energy in the cosmological frame scales
as $E_{\rm p} \propto [\gamma(\theta_{\rm obs} -\theta_0)]^{-2}$ 
\citep[e.g.,][]{Granot02}. This
also implies that when seen off-axis $E_{\rm p}$ will fall away from
the Amati relation \citep{Amati02,L-RR-R02}, $E_{\rm p} \propto
E_{\gamma,{\rm iso}}^{1/2}$, by a factor of $\gamma(\theta_{\rm obs}
-\theta_0)$ (Fig. \ref{fig3}). The low $E_{\gamma,{\rm iso}}$ of
GRB 031203 implies\footnote{This follows from the scaling 
$E_{\gamma,{\rm iso}} \propto [\gamma(\theta_{\rm obs}-\theta_0)]^{-6}$
for $\gamma(\theta_{\rm obs}-\theta_0)\gtrsim 1$, assuming that the 
energy radiated in the prompt emission, 
$E_\gamma\approx(\theta_0^2/2)E_{\rm\gamma,iso}(\theta_{\rm obs}<\theta_0)$, 
is comparable to $E_{\rm jet}$.}
\begin{equation}\label{theta_0}
\theta_0 = 3.8^\circ 
\left({E_{\gamma,{\rm iso}} \over 10^{50}\;
{\rm erg}}\right)^{-1/8} 
\left(\frac{E_{\rm jet}}{3\times 10^{50}\;{\rm erg}}\right)^{1/8}
\left(\frac{\gamma\Upsilon}{50}\right)^{-3/4}\ ,
\end{equation}
where $E_{\rm jet}$ is the kinetic energy of the jet, and
$\Upsilon=\theta_{\rm obs}/\theta_0-1$. The fiducial values in Eq.
\ref{theta_0} were chosen to match those of GRB 031203, which were
either observed ($E_{\gamma,{\rm iso}}\sim 10^{50}\;$erg) or inferred
from the fit to its afterglow ($\theta_0\sim 3^\circ-5^\circ$, $E_{\rm
jet}\sim 3\times 10^{50}\;$erg, $\Upsilon\approx 1$), and they imply
$\gamma\sim 50$. Eq. \ref{theta_0} gives
\begin{equation}
\gamma(\theta_{\rm obs} -\theta_0) =
3.3\left({E_{\gamma,{\rm iso}} \over 10^{50}\;
{\rm erg}}\right)^{-1/8} 
\left(\frac{E_{\rm jet}}{3\times 10^{50}\;{\rm erg}}\right)^{1/8}
\left(\frac{\gamma\Upsilon}{50}\right)^{1/4}\ ,
\end{equation}
which implies more typical values of $E_{\rm p} \sim 2\;$MeV (given
the observed value $E_{\rm p} \sim 190\;$keV) and $E_{\gamma,{\rm
iso}} \sim 10^{53}\;$ergs when observed on-axis
(Fig. \ref{fig3}). These values fall somewhat above the Amati
relation, but this is not alarming given that a reasonable fraction of
BATSE bursts are also not consistent with this empirical
law\footnote{although this conclusion is debated \citep[see
e.g.][]{bos05}.} \citep[e.g.][]{NP04}.

These results are applicable in the present context provided only that
one further condition is satisfied, namely, that the (on axis) jetted
outflow be optically thin to high-energy photons \citep[e.g.][]{LS01}.
For a burst with $E_{\rm p} \sim 2$ MeV, $\gamma$ must exceed $\sim
50$.

We consider the required value of $\gamma\sim 50$ and an inferred core
value of $E_{\rm p}\sim 2$ MeV to be reasonable for a jet viewed
outside of the core. Close to the rotation axis $\gamma$ may be high
while near its edge there will likely be an increasing degree of
entrainment with a corresponding decrease in $\gamma$
\citep{ZWH04}. Moreover, in the internal shock model, $E_{\rm
p}\propto\gamma^{-2}$ \citep[e.g.][]{RL02} so that for most lines of
sight within the jet aperture, where $\gamma$ is slightly higher than
in the edges, an observer would naturally detect bursts with lower
values of $E_{\rm p}$. Off-axis observers, on the other hand, see
mainly the edge of the jet where $\gamma$ is lower than in the axis
and would thus tend to infer higher (on-axis) values of $E_{\rm p}$.

Another possibility is that the jet does not have sharp edges but
wings of lower energy and Lorentz factor that extend to large
$\theta$. Such a picture of the jet was suggested by \citet{WES99} and
is consistent with the relativistic studies of the collapsar model by
\citet{R-RCR02} and \citet{ZWH04}.  GRB 031203 would then be produced
by the interaction of relativistic material moving in our
direction with the circumstellar medium -- the wind of the
pre-explosive star.  Unfortunately it is difficult, in the simplest
version of this model, to account for the prompt emission in GRB
031203. If one is restricted to producing the burst by an external
shock interaction using a geometrically thin blast wave, the observed
duration and hardness are incompatible. Details of this model and
attempts to extend it will be discussed elsewhere.

\section{Conclusion}

The characteristic energy scale for common GRBs has been debated for a
long time, in particular the question of whether all GRBs are, in some
sense, a standard explosion with a nearly constant energy. The GRB
community has vacillated between initial claims that the GRB intrinsic
luminosity distribution was very narrow \citep{Horack92}, to
discounting all standard candle claims, to accepting a standard total
GRB energy of $\sim 10^{51}$ ergs \citep{Frail01}, to diversifying
GRBs into ``normal'' and ``sub-energetic'' classes (S04).

The recent discovery of the faint GRB 031203 has been argued to
support the existence of at least two classes of GRB/SN Ib/c events
based on different amounts of energy released during the initial
explosion. In this {\it Letter}, we have examined two possible
interpretations of the observations of GRB 031203 based upon the
premise that it was either an {\it ordinary} GRB observed off-axis or
an intrinsically weak, nearly isotropic explosion. We conclude that
the observations, especially the slow initial decline rates seen in
the X-ray afterglow, are more consistent with an off-axis model in
which GRB 031203 was a much more powerful GRB seen at an angle of
about two times the opening angle of the central jet\footnote{This
conclusion is also supported by a statistical argument for the number
of observed low redshift GRBs \citep{Guetta04}.}. Early and detailed
X-ray observations of GRB afterglows would provide more stringent
constraints on the jet geometry and energetics.

\acknowledgments This work is supported by IAS and NASA through a
Chandra Fellowship award PF3-40028 (ERR) and by the DoE under contract
DE-AC03-76SF00515 (JG). The authors acknowledge benefits from
collaboration within the RTN "GRBs: An Enigma and a Tool". At UCSC,
this research was supported by the NSF (AST 02-06111) and NASA
(NAG5-12036).


\begin{thebibliography}{}
 
\bibitem[Amati et al.(2002)]{Amati02} 
Amati, L.  et al. 2002, A\&A, 390, 81
 
\bibitem[Bosnjak et al.(2005)]{bos05} Bosnjak, Z., Celotti, A., Longo
F., \& Barbiellini, G. 2005, submitted to MNRAS (astro-ph/0502185)

\bibitem[Cobb et al.(2004)]{cobb04} Cobb, B.~E. et al. 2004, ApJ, 608,
L93

\bibitem[Galama et al.(1998)]{Galama98} 
Galama, T. J. et al. 1998, Nature, 395, 670

\bibitem[Frail et al.(2001)]{Frail01} 
Frail, D.~A. et~al. 2001, ApJ, 562, L55

\bibitem[Frail et al.(2003)]{Frail03} 
Frail, D. A., Kulkarni, S. R., Berger, E. \& Wieringa,
M. H. 2003, AJ, 125, 2299

\bibitem[Gal-Yam et al.(2004)]{galyam04} Gal-Yam, A. et al. 2004, ApJ,
609, L59

\bibitem[Ghirlanda, Ghisellini \& Lazzati(2004)]{GGL04} 
Ghirlanda, G., Ghisellini, G., \& Lazzati, D. 2004, ApJ in
press, astro-ph/0405602

\bibitem[Granot et al.(2001)]{Granot01}
Granot, J, Miller, M., Piran, T., Suen, W.~M., \& Hughes, P.~A.
2001, in Gamma-Ray Bursts in the Afterglow Era, ed. E. Costa, F.
Frontera, \& J. Hjorth (Berlin: Springer), 312

\bibitem[Granot et al.(2002)]{Granot02} 
Granot, J., Panaitescu, A., Kumar, P., \& Woosley,
S.~E.\ 2002, ApJ, 570, L61

\bibitem[Granot \& Kumar(2003)]{GK03} 
Granot, J., \& Kumar, P. 2003, ApJ, 591, 1086

\bibitem[Granot \& Ramirez-Ruiz(2004)]{GR04} Granot, J., \&
Ramirez-Ruiz, E. 2004, ApJ, 609, L9

\bibitem[Granot, Ramirez-Ruiz \& Perna(2005)]{GR-RP05} 
Granot, J., Ramirez-Ruiz, E., \& Perna, submitted to ApJ (astro-ph/0502300)

\bibitem[Guetta et al.(2004)]{Guetta04}
Guetta, D., Perna, R., Stella, L., \& Vietri, M. 2004, ApJ, 615, L73

\bibitem[Horack et al.(1992)]{Horack92} 
Horack, J.~M. et al. 1992, ApJ, 426, L5

\bibitem[Kouveliotou et al.(2004)]{Kouveliotou04} 
Kouveliotou, C. et al. 2004, ApJ, 608, 872

\bibitem[Kulkarni et al.(1998)]{Kulkarni98} 
Kulkarni et al. 1998, Nature, 395, 663

\bibitem[Lithwick \& Sari(2001)]{LS01} 
Lithwick, Y., Sari, R. 2001, ApJ, 555, 540

\bibitem[Lloyd-Ronning \& Ramirez-Ruiz(2002)]{L-RR-R02} 
Lloyd-Ronning, N., \& Ramirez-Ruiz, E. 2002, ApJ, 576, 101

\bibitem[Malesani et al.(2004)]{Malesani04} Malesani, J. et al. 2004,
ApJ, 609, L5

\bibitem[Nakar \& Piran(2004)]{NP04}
Nakar, E., \& Piran, T. 2004, preprint (astro-ph/0412232)

\bibitem[Prochaska et al.(2004)]{Prochaska04} 
Prochaska, J.~X. et al. 2004, ApJ, 611, 200

\bibitem[Ramirez-Ruiz, Celotti \& Rees(2002)]{R-RCR02} 
Ramirez-Ruiz, E., Celotti, A., \& Rees, M. J. 2002, MNRAS, 337, 1349

\bibitem[Ramirez-Ruiz \& Lloyd-Ronning(2002)]{RL02} Ramirez-Ruiz, E.,
Lloyd-Ronning, N.~M. 2002, New Astron., 7, 197

\bibitem[Ramirez-Ruiz \& Madau(2004)]{R-RM04} Ramirez-Ruiz, E., \&
Madau, E. 2004, ApJ, 608, L89

\bibitem[Sazonov et al.(2004)]{SLS04} 
Sazonov, S.~Y., Lutovinov, A.~A.  \& Sunyaev, R.~A.,
2004, Nature, 430, 646

\bibitem[Sari, Piran \& Narayan(1998)]{SPN98} 
Sari, R., Piran, T., \& Narayan, R.\ 1998, \apjl, 497, L17 

\bibitem[Soderberg et al.(2004)]{Soderberg04} 
Soderberg, A.~M. et~al. 2004, Nature, 430, 648
 
\bibitem[Thomsen et al.(2004)]{Thomsen04} 
Thomsen, B. et al. 2004, A\&A, 419, L21

\bibitem[Watson et al.(2004a)]{Watson04a} 
Watson, D. et al. 2004a, ApJ, 605, L101

\bibitem[Watson et al.(2004b)]{Watson04b} Watson, D. et al. 2004b,
A\&A, 425, L33

\bibitem[Waxman(2004)]{Waxman04}
Waxman, E. 2004, ApJ, 605, L97

\bibitem[Woods \& Loeb(1999)]{WL99}
Woods, E., \& Loeb, A. 1999, ApJ, 523, 187

\bibitem[Woosley, Eastman \& Schmidt(1999)]{WES99} 
Woosley, S., Eastman, R., \& Schmidt, B. 1999, ApJ, 516,
788

\bibitem[Yamazaki et al.(2002)]{YIN02}
Yamazaki, R., Ioka, K, \& Nakamura, T. 2002, ApJ, 571, L31

\bibitem[Zhang, Woosley \& Heger(2004)]{ZWH04} 
Zhang, W., Woosley, S. E., \& Heger, A. 2004, ApJ, 608,
365

\end{thebibliography}
\end{document}